\documentclass[12pt]{iopart}
\usepackage{graphicx}
\usepackage{amssymb}
\usepackage{color}
\usepackage{ulem}
\usepackage{braket}
\usepackage{comment}
\usepackage{soul}
\usepackage{cite}
\DeclareMathAlphabet{\mathantt}{OT1}{antt}{li}{it}

\newcommand{\Sch}{{Schr\"{o}dinger }}

\begin{document}

\title{Cavity polaritons with Rydberg blockade and long-range interactions}
\author{Marina Litinskaya, Edoardo Tignone, and Guido Pupillo}
\address{icFRC, IPCMS (UMR 7504) and ISIS (UMR 7006), Universit$\acute{\rm e}$ de Strasbourg and CNRS, 67000 Strasbourg, France}

\begin{abstract}
We study interactions between polaritons, arising when photons strongly couple to collective excitations in an array of two-level atoms trapped in an optical lattice inside a cavity. We consider two types of interactions between atoms: Dipolar forces and atomic saturability, which ranges from hard-core repulsion to Rydberg blockade. We show that, in spite of the underlying repulsion in the subsystem of atomic excitations, saturability induces a broadband bunching of photons for two-polariton scattering states. We interpret this bunching as a result of interference, and trace it back to the mismatch of the quantization volumes for atomic excitations and photons. We examine also bound bipolaritonic states: These include states created by dipolar forces, as well as a gap bipolariton, which forms solely due to saturability effects in the atomic transition. Both types of bound states exhibit strong bunching in the photonic component. We discuss the dependence of bunching on experimentally relevant parameters.
\end{abstract}




\section{Introduction}

There is a growing interest in studying the effects of exciton-polariton physics in solid state systems \cite{Ciuti,Deng,Byrnes,Deveaud,Laussy2012}. Nonlinearities in semiconductor microcavities have already led to the experimental demonstration of fundamental phenomena such as  superfluidity \cite{Amo} as well as Bose-Einstein condensation \cite{Kasprzak,Balili,Menard} of exciton-polaritons (or simply polaritons). In organic materials, Bose-Einstein condensates (BECs) of cavity polaritons have been recently obtained even at room temperature \cite{Plumhof,Daskalakis}. 
For inorganic semiconductors, the capability of fabricating complex microcavities allows for a high degree of tunability of experimental parameters \cite{Besga}, which paves the way towards the quantum simulation of condensed matter phenomena with polariton BECs \cite{Wertz,Tanese,Jacqmin,Baboux}.  
In the context of atomic physics and quantum optics, due to their strong dipole-dipole interactions \cite{Galla}, Rydberg-excited atoms \cite{book0} are a favorable atomic platform to induce photon-photon interactions \cite{Weatherill,Hofmann}. Thanks to the dipole blockade \cite{Grangier,Urban,Pillet,Zoller,Ott,Ebert}, polaritons in gases of Rydberg-excited atoms under conditions of electromagnetically induced transparency (EIT) \cite{eit,Mohapatra,RSV} provide an efficient way to induce strong optical nonlinearities in cavities \cite{Grankin0,Grankin,Boddeda} as well as in free space \cite{Alexey,Gauguet,Tresp,Moos,Gunter}, store and manipulate photons \cite{Szwer,Molmer}, realize single-photon phase-shifts \cite{Tiarks2015} and transistors \cite{Tiarks2014,Gorniaczyk2014,Durr}, and generate nonclassical states of light \cite{Li,Otterbach,Kuzmich,Yao,Molmer2012,Vuletic,Vuletic1}. In particular, both dispersive \cite{Vuletic} and dissipative \cite{Vuletic1} quantum nonlinearities have been realized in Rydberg gases under EIT leading to effective photon-photon attraction and repulsion, respectively. It was shown in theoretical studies \cite{Buchler,Maghrebi} that few-photon attraction in these systems can be described in terms of formation of bound bipolariton states in effective potentials.

\begin{figure}[t!]
\centering
\includegraphics[width=\columnwidth]{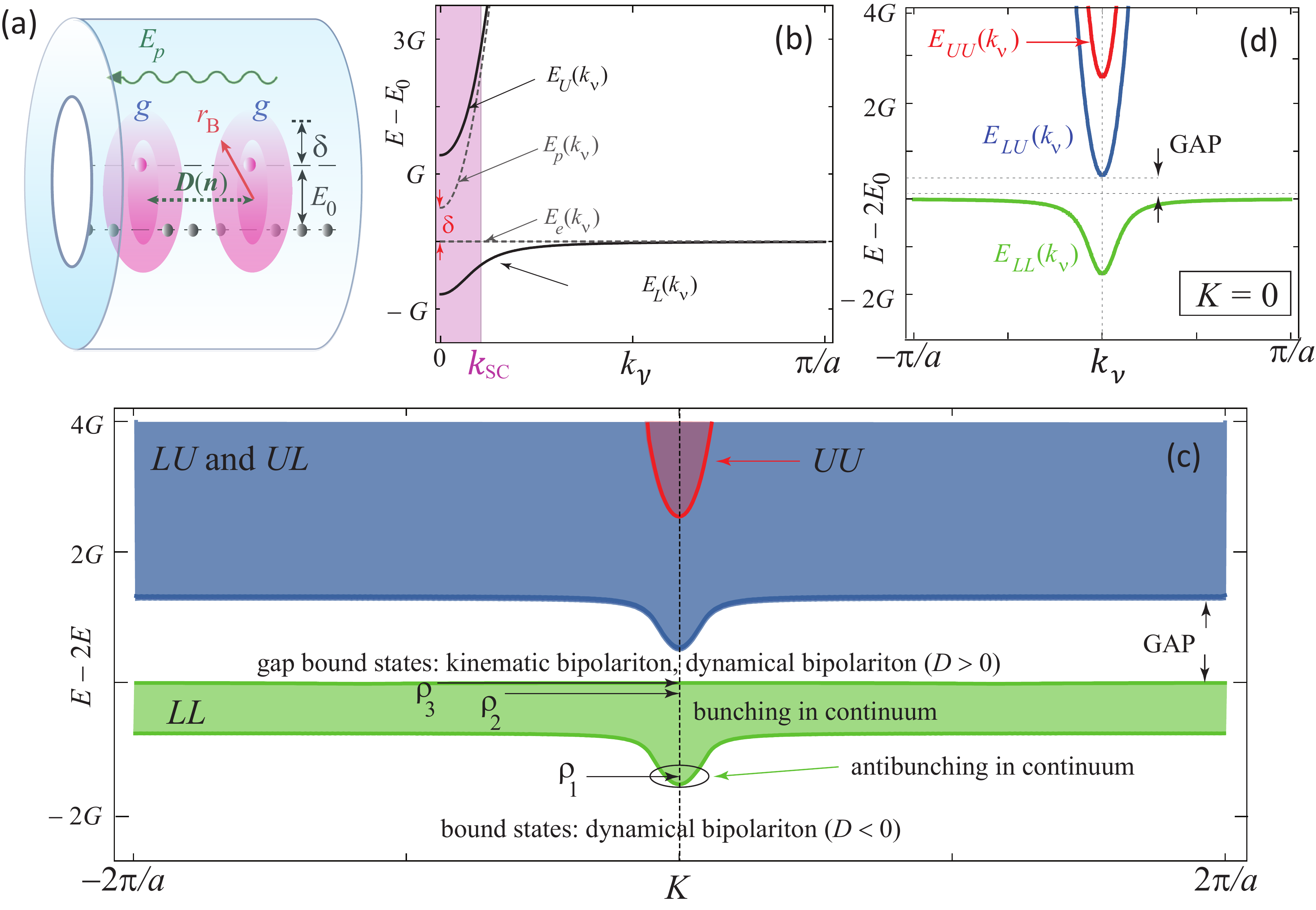}
\caption{(Color online) (a) Sketch of the model setup: An array of evenly spaced two-level atoms is placed within a cavity in a one-dimensional configuration. The atom-photon coupling is $g$, while $E_p$ and $E_0$ are the energy of the cavity mode and of the atomic transition, respectively, with $\delta=E_p-E_0$ the detuning. Atoms can interact with each other both via a direct (dipole-dipole or van der Waals) interaction $D$, and/or via the Rydberg blockade mechanism. The blockade radius $r_B$ characterizes the volume around each atomic excitation where a second excitation is inhibited (pink shaded region). (b) The dispersions of the lower [$E_L(k_\nu)$] and of the upper [$E_U(k_\nu)$] polaritons of equation (\ref{polaritons}) are indicated by solid lines, and compared to those of the bare photon [$E_p(k_\nu)$] and exciton [$E_{e}(k_\nu)$] (dashed lines). The wave vector $k_{\rm SC}$ defines the size of the strong coupling region; $a$ is the lattice constant. (c) Two-polariton bands (no interactions) as functions of total momentum $K$, together with a summary of effects that result from finite interactions. Labels $\rho_1$, $\rho_2$ and $\rho_3$ mark the states shown below in figure~\ref{f-compare wave functions}. (d) Two-polariton states for $K = 0$ as function of the relative momentum $k_\nu$, which is another good quantum number in absence of polariton-polariton interactions. Here, $E_{LL}(k_\nu)=2E_L(k_\nu)$, $E_{LU}(k_\nu)=E_L(k_\nu)+E_U(k_\nu)$, and $E_{UU}(k_\nu)=2E_U(k_\nu)$}.
\vskip -0.2cm 
\label{f-spaceholder1}
\end{figure}

In this work we continue to study analytically and numerically a theoretical model \cite{D2-paper} that shares features both with the solid-state and the atomic-physics polariton setups: It describes interactions between two cavity photons that are coupled to collective excitations in an extended, ordered ensemble of two-level atoms (or {\it excitons}) [figure \ref{f-spaceholder1}(a)] confined to a one-dimensional (1D) geometry. In the model, the linear size of the atomic array largely exceeds the photon wave length. If the coupling between photons and excitons is much stronger than the losses, the eigenmodes are coherent superpositions of excitons and photons \cite{book}; see figure~\ref{f-spaceholder1}(b). For these states -- which are known as polaritons -- the atomic absorption is eliminated by the ordering of the atoms (see Discussion section for details). This distinguishes ``direct" (i.e., two-level) polaritons from those generally used in three-level configurations within a disordered Rydberg atomic gas under the condition of EIT, which eliminates absorption \cite{Vuletic,Vuletic1}.

Figure \ref{f-spaceholder1}(c) shows two-polariton states as function of total momentum $K$, which is a good quantum number in a translationally invariant system. The detuning $\delta$ [see figure \ref{f-spaceholder1}(b)] between the cavity mode and the exciton is chosen to be positive. The spectrum consists of four bands: lower-lower (LL), and overlapping lower-upper (LU), upper-lower (UL), and upper-upper (UU) bands. When $\delta \to 0$, the minimum of LU-UL continuum touches the top of LL-band, and the gap closes at $K=0$. Due to the lattice geometry, all bands have a finite width. The LL-band is relatively narrow (of the order of twice the collective light-matter coupling constant $G$, see below), while the LU-, UL- and UU-bands spread till much higher energies, so that only their lowest states are relevant to the polariton dynamics. Finally, in the absence of polariton-polariton interactions, the relative wave vector $k_\nu$ also is a good quantum number. Figure \ref{f-spaceholder1}(d) shows two-polariton bands as function of $k_\nu$ for $K=0$ [i.e. along the dashed line cut in figure \ref{f-spaceholder1}(c)]. For $K\neq 0$, two minima are obtained for the LL-band at $k_\nu = \pm K/2$, and the LU- and UL-bands split, with minima at $k_\nu = K/2$ (LU) and $k_\nu = - K/2$ (UL).

Polaritons interact with each other via the interactions in the atomic subsystem, which are of two types. The first type, often referred to as {\it kinematic interaction}, is a result of atomic saturability, and reflects the fact that one atom can accommodate at most one excitation, due to the intrinsic non-linearity of the atomic spectrum. This corresponds to a ``hard-core" repulsion between the excitations, e.g., for atoms trapped in optical lattice (see below). In an extreme form, the kinematic interaction can also be used to describe the Rydberg blockade, which occurs when a local atomic excitation shifts the energy levels of neighboring atoms, thus inhibiting the absorption of a second photon within a given blockade radius~\cite{Zoller,Urban,Gauguet}. The second type of interaction -- the {\it dynamical interaction} -- is the usual direct long-range interaction between Rydberg excited atoms, which can be of the dipole-dipole or van der Waals type \cite{Molmer}.\\

The paper contains the following results, which are schematically summarized in the total momentum vs. energy plane in figure~\ref{f-spaceholder1}(c). {\it First}, we discuss the effects of the kinematic interaction. In \cite{D2-paper} we have shown that it induces correlations between photons, which, for typical atomic parameters, are visible in the continuum of polariton-polariton states. Here we generalize these results for finite blockade radii. As opposed to the narrow bound-state resonances of Rydberg polaritons under EIT conditions \cite{Vuletic}, this effect occurs in a comparatively wide  frequency range, which can be of the order of tens of GHz. Depending on the interplay between the exciton-photon coupling, the strength of exciton-exciton interaction and the photon frequency, both bunching and antibunching can be observed in the photonic component. The two-photon bunching is enhanced for larger radii $r_B$ of the Rydberg-blockade interaction. {\it Second}, we study the role of the dynamical interaction. We demonstrate that the dynamical interaction does not affect photon bunching in the continuum. Its main effect, in the case of attraction, is the appearance of bipolaritons -- bound two-polariton states that appear below the two-polariton continuum. These are narrow resonances that also realise strong bunching in the photonic component if $r_B$ is small; in contrast, for large $r_B$ the photons are essentially bound to the excitons, which are now separated by the blockade distance $2r_B+a$ ($a$ being the lattice spacing; $r_B/a$ is an integer), and thus— antibunch. {\it Third}, we consider polaritonic spectra with positive detuning $\delta$ between the cavity mode and the exciton [see figure~\ref{f-spaceholder1}(b)], when the two-polariton spectrum has a gap between the lowest-energy and next-to-lowest-energy bands. We show that bound bipolaritonic states can form also within this gap, due to either repulsive dynamical, or kinematic interaction. These bound states exhibit a photonic bunching, which is stronger, the larger $r_B$.

In our model the atoms are regularly spaced from each other and trapped in a 1D configuration inside a cavity. A brief discussion of the experimental conditions and challenges for the experimental realisation of this model for a generic choice of Rydberg states is presented in Section ~\ref{s-conclusions}. We note, however, that, neglecting dissipation, a  configuration similar to that discussed here has been already realised for cold Sr atoms trapped in an optical lattice  \cite{Sr} within a hollow-core photon crystal fiber (HC-PCFs) \cite{Na,Blatt,Bajcsy}, and excited on the D2-line transition; more generally, in the last few years a great effort has been expended on 1D nonlinear nanophotonic platforms \cite{ Zoubi, Douglas,Hafezi, Hafezi11, Kimble, Yu,Liao2015, Goban, Nayak, Vetsch, Sayrin,Rauschenbeutel,Mitsch, Reitz,Faggiani}.
Recent experiments have also demonstrated the loading  of HC-PCFs with atomic vapors at room temperature \cite{Slepkov,Venkataraman,Vogl2014} and in Rydberg states \cite{Epple}. In different contexts, the direct two-level excitations of Rydberg states considered here have also been proposed \cite{Gross3,Gross4} and demonstrated~\cite{Biedermann,Tong,Deiglmayr,Manthey}.

The paper is organized as follows. In Section~\ref{s-model} we introduce our model and basic notations, and derive the equations describing interacting polaritons. In Section~\ref{s-KI} we study the kinematic interaction, assuming that the dynamical interaction is switched off. We derive an exact analytical solution for excitons in the presence of a blockade sphere with finite radius $r_B$, and use these results to interpret the role of kinematic interactions for polaritons. In particular, we show that the two-photon correlations are a result of interference between different photonic components coupled via the excitonic subsystem, which samples a reduced quantization volume. The role of the dynamical interaction and the formation of low-energy dynamical bipolaritons are discussed in Section~\ref{s-D}. The gap states (dynamical and kinematic bipolaritons) are discussed in Section~\ref{s-gap}. Finally, we discuss the experimental realizability of the obtained results and present an outlook in Section~\ref{s-conclusions}.


\section{Model and notations}
\label{s-model}

The Hamiltonian describing an ordered array of $N$ atoms coupled to the photon field of a cavity in one dimension reads
\begin{equation}\label{Hamiltonian}
\begin{array}{c}
\fl H = E_0 \sum\limits_s P_s^{\dagger} P_s + t \sum\limits_{s} \left(P_s^{\dagger}
P_{s+1} + P_s^{\dagger} P_{s-1} \right) + \frac{1}{2}\sum\limits_{s,p} D(s-p) P_s^{\dagger} P_p^{\dagger} P_s P_p + \\

\\

\fl + \sum\limits_{q_\nu} E_p(q_\nu) a^{\dagger}(q_\nu) a(q_\nu) 
+ g \sum\limits_{s,q_\nu} \left[P_s^{\dagger} a(q_\nu) e^{iq_\nu s}+P_s a^{\dagger}(q_\nu)
e^{-iq_\nu s}\right].\\
\end{array}
\end{equation}

Here $P_s$, $P_s^{\dagger}$ destroy and create an atomic excitation at site $s$, respectively, while $a(q_\nu)$, $a^{\dagger}(q_\nu)$ destroy and create a photon with a wave vector $q_\nu = 2\pi\nu/(Na)$ directed along the cavity axis ($\nu$ is an integer with $|\nu|\leq N/2$, $N$ is the number of atoms in the cavity, and $a$ is the lattice constant). Furthermore,  $E_p(q_\nu) = c\hbar \sqrt{q_\nu^2 + q_\perp^2}$ is the energy of a cavity photon, with $q_\perp$ the quantized transversal momentum, $c$ the speed of light. The atomic subsystem is described by the transition energy $E_0$ between the ground state and an excited (Rydberg) state; $t$ is the dipole-dipole induced excitation hopping constant in the nearest-neighbor approximation, while $D(s-p)$ accounts for the long-range interaction between sites s and p (dipole-dipole or van der Waals); $g$ is the atom-light coupling constant. In the absence of a long-range $D$-term, the atomic part of the Hamiltonian is diagonalized by Frenkel exciton operators \cite{book} $P(q_\nu) = \frac{1}{\sqrt{N}} \sum_n P_n e^{-iq_\nu n}$, which describe extended collective atomic excitations resulting from hopping. The effects of exciton hopping, however, are usually minor in the cavity configuration, as the exciton dispersion is much smaller than that of polaritons originating from light-matter coupling.\\

In the following, we solve the Schr\"odinger equation in the subspace of two excitations, where a wave function has the general form
\begin{equation}\label{Psi}
| \Psi \rangle = \sum_{nm} \left\{ \frac{A_{nm}}{\sqrt{2}} \ket{a^\dag_n a^\dag_m} + B_{nm} \ket{a^\dag_n P^\dag_m} + \frac{C_{nm}}{\sqrt{2}} \ket{P^\dag_n P^\dag_m} \right\}.
\end{equation}
Here, $a_n = \frac{1}{\sqrt{N}}\sum_{q_\nu} a(q_\nu) e^{i q_\nu n}$, and $A_{nm}, B_{nm}=B_{nm}^S + B_{nm}^A$ and $C_{nm}$ are the amplitudes of finding two bare particles at sites $n,m$. The superscripts $``S",``A"$ stay for symmetric and antisymmetric; the amplitudes $A_{nm}$ and $C_{nm}$ are always symmetric.

Each atom can absorb at most one photon. This induces correlations between one-particle eigenstates, which are known as {\it kinematic interactions} \cite{book}. While in natural solids the kinematic interaction is usually a very weak effect, in cold atomic systems the latter can be comparatively large \cite{D2-paper}. In addition, in Rydberg gases an excitation of one atom may suppress the excitation probability for neighboring atoms via a shift of their energy levels induced by the dynamical interaction $D$ \cite{Zoller,Urban,Saffman}. The resulting blockade radius is usually of the order of several $\mu$m and may largely exceed the interatomic spacing $a$, which, for atoms trapped in an optical lattice, is usually of the order of several hundred nm. In the following, we account for the Rydberg blockade mechanism by introducing the blockade radius $r_B$ and assuming that double excitations are suppressed within a region of length $2r_B+a$; $r_B = 0$ corresponds to the ``usual" kinematic interaction resulting in an on-site hard-core-type constraint, while we term the finite $r_B$ due to the Rydberg blockade mechanism as ``extended kinematic interaction". The kinematic interaction can be accounted for in our analytical treatment by removing the states $\ket{P_n P_m}$ with $|n-m| < 2(r_B/a)+1$ from the total basis set, similar to the case of the hard-rod Tonks gas in free space \cite{Tonks}. Therefore, we can solve the \Sch equation for the Hamiltonian (\ref{Hamiltonian}) by eliminating the basis states $\ket{P_n P_m}$, which is achieved by multiplying the corresponding amplitudes by $[1-\theta(n-m)]$, where $\theta(x)$ is the step function with $\theta(x) = 1$ if $|x| \leq r_B/a$ and zero otherwise. For convenience, we then set
\begin{equation}\label{constraint}
\theta(n-m)C_{nm} = 0.
\end{equation}

To find the eigenstates, $(i)$ we symmetrize the equations for the amplitudes, $(ii)$ Fourier transform the result,  and $(iii)$ rewrite the obtained equations in terms of total and relative wave vectors: $K_{\nu^{'}} = q_{\nu_1}+q_{\nu_2}$ and $k_{\nu} = (q_{\nu_1} - q_{\nu_2})/2$, with $q_{\nu_1}$ and $q_{\nu_2}$ being the wave vectors of two bare excitations (for example, one exciton and one photon). The total wave vector $K_{\nu^{'}}$ is the quantum number describing the two-excitation spectra, while $k_{\nu}$ is a simple label. The derivation for the general case is cumbersome and is presented in the Appendix. Below we present analytical results for $K_{\nu^{'}} = 0$; the numerical results for $K_{\nu^{'}} \neq 0$ are shown later in figure \ref{f-bunching}(b,c). For $K_{\nu^{'}} = 0$, $q_{\nu_2} = -q_{\nu_1}$, and the relative wave vector reads:

\begin{equation}\label{k_nu}
k_\nu = \frac{2\pi\nu}{Na}, \hskip 1cm \nu = -\frac{N}{2}+1, ... , \frac{N}{2}.
\end{equation}
The two-particle amplitudes are found to obey a simple set of equations:

\begin{equation}\label{Schroedinger-2}
\begin{array}{l}
E_\rho A_\rho(k_\nu) = 2E_{p}(k_\nu) A_\rho(k_\nu) + G\sqrt{2} B_\rho(k_\nu),\\

\\

E_\rho B_\rho(k_\nu) = \left[ E_{p}(k_\nu) + E_{e}(k_\nu) \right] B_\rho(k_\nu) + G\sqrt{2} [A_\rho(k_\nu) + C_\rho(k_\nu)],\\

\\

E_\rho C_\rho(k_\nu) = 2E_{e}(k_\nu) C_\rho(k_\nu) + G\sqrt{2} B_\rho(k_\nu) + S_\rho(k_\nu),\\
\end{array}
\end{equation}
where $E_p(k_\nu)$ and $E_e(k_\nu) = E_0 + 2t\cos ak_\nu$ are the energies of a photon and an exciton (in $k$-representation), respectively, and $G = g\sqrt{N}$ is the collective atom-photon coupling constant. The index $\rho$ is introduced to enumerate two-polariton states, and

\begin{equation}\label{S(k)}
\fl S_\rho(k_\nu) = \frac{1}{N} \sum\limits_{q_\nu} \biggl\{  D(k_\nu-q_\nu) C_\rho(q_\nu) - \theta(k_\nu-q_\nu) G\sqrt{2} B_\rho(q_\nu) - \theta(k_\nu-q_\nu)\ 4 t \cos aq_\nu\  C_\rho(q_\nu)\biggr\}\\
\end{equation}
accounts for polariton-polariton scattering due to both the dynamical and kinematic types of interactions.\\

In the absence of all interactions, i.e.~with $S_\rho(k_\nu) \equiv 0$, this system of equations (\ref{Schroedinger-2}) has non-trivial solutions when
\begin{equation}
\Delta (E,k_\nu) \equiv [E - 2E_L(k_\nu)][E - E_L(k_\nu)-E_U(k_\nu)][E - 2E_U(k_\nu)] = 0.
\end{equation}

Here
\begin{equation}\label{polaritons}
E_{L,U}(k_\nu) = \frac{1}{2} \left\{ E_e(k_\nu) + E_p(k_\nu) \mp \sqrt{(E_e(k_\nu)-E_p(k_\nu))^2 + 4 G^2} \right\}
\end{equation}
denote the energies of the lower (with ``$-$") and the upper (with ``+") non-interacting polaritons shown in figure ~\ref{f-spaceholder1}(b).
The two-particle eigenenergies in the non-interacting limit are $E_{ij}(k_\nu) = E_{i}(k_\nu) + E_{j}(k_\nu)$ with $i,j \in L,U$ [see figure \ref{f-spaceholder1}(c)].

The term $S_\rho(k_\nu) \neq 0$ introduces interactions between free two-polariton states. We have shown in \cite{D2-paper} that for the kinematic interaction with $r_B = 0$ (and  $D = 0$) the interacting eigenstates are described by a set of effective wave vectors, which with the increase of $\rho$ make a gradual transition from free-like wave vectors (labeled as $k_\nu$) to a new wave vector set, corresponding to interacting excitons. This latter excitonic set accounts for the excluded blockade volume resulting from exciton-exciton correlations. In particular, in \cite{D2-paper} we have shown that the repulsion in the excitonic subsystem can result in an effective attraction (bunching) in the photonic subsystem. Below we generalize our results for arbitrary sizes $(2r_B+a)$ of the excluded Rydberg blockade volume, and discuss the role of long-range dynamical interactions for photon-photon correlations.

\section{Kinematic interaction}
\label{s-KI}

In this section we set $D=0$ and concentrate on the effect produced by the extended kinematic interaction. The role of the dynamical interaction is discussed in Section~\ref{s-D}.

\subsection{Kinematic interaction of excitons}

We start with the exact solution of the \Sch equation for two bare excitons interacting via extended kinematic interaction, by considering only the first two terms of the Hamiltonian (\ref{Hamiltonian}) and the last term of the wave function (\ref{Psi}). Let $n = |n_1-n_2|$ denote the relative distance between two atomic excitations located at sites $n_1$ and $n_2$, and the index $\mu$ enumerate the eigenstates of the resulting Schr\"odinger equation. Using the hard-core constraint (\ref{constraint}), we write the eigenstate equation as

\begin{equation}\label{kinematic_ex}
\fl E^{(ex)}_\mu C^{(ex)}_\mu(n) = [1 - \theta(n)] \biggl\{ 2E_0 C^{(ex)}_\mu(n) + 2t[ C^{(ex)}_\mu(n+1) + C^{(ex)}_\mu(n-1)] \biggr\}.
\end{equation}
One can check by direct substitution that the normalized set of excitonic amplitudes $C_\mu^{(ex)}(n)$ that satisfy this equation is given by (we remind that $r_B/a$ is an integer)
\begin{equation}\label{Cex(n)}
C_\mu^{(ex)}(n) \equiv  g_n(\mu) = \frac{\sqrt{2}[1 - \theta(n)]}{\sqrt{N - 2(r_B/a)}}\ \sin \kappa_\mu [|n|-(r_B/a)],
\end{equation}
where $\kappa_\mu$ represent a ``new" wave vector set
\begin{equation}\label{k renormalized}
\kappa_\mu = \frac{2\pi |\mu| }{Na - 2r_B}, \hskip 1cm \mu = -\frac{(N-1)}{2}+\frac{r_B}{a},..., \frac{(N-1)}{2} - \frac{r_B}{a},
\end{equation}
with eigenenergies given by
\begin{equation}\label{Eex}
E^{(ex)}_\mu = 2E_0 + 4t \ \cos a \kappa_\mu.
\end{equation}

The basis functions $g_n(\mu)$ form an orthonormal set in the spaces of $|n|$ and $|\mu|$:
\begin{equation}
\label{eq:gprops}
\fl \sum_n g_n(\mu_1)g_n(\mu_2)=\delta_{|\mu_1|,|\mu_2|}, \hskip 1cm
\sum_\mu {g_{n_1}}(\mu){g_{n_2}}(\mu)=[1-\theta(n_1)]\delta_{|n_1|,|n_2|}.
\end{equation}

The equations above have the following interpretation: While the non-interacting excitons are described by the ``original wave vector set"~(\ref{k_nu}), excitons in the presence of extended kinematic interaction are classified by the ``reduced wave vector set"~(\ref{k renormalized}). The latter has two distinct features: (i) The normalization volume takes into account the Rydberg sphere, and thus reads $Na - 2r_B$ instead of $Na$, and (ii) the state index $\mu$ takes peculiar {\it half-integer} values, so that the reduced wave vectors $\kappa_\mu$ fall between the values of the original wave vectors $k_\nu$. As a consequence, the amplitudes $C_\mu^{(ex)}(k)$ do not have poles (that is, their denominators never vanish); instead they have enhanced contributions from the components of $k_\nu \approx \kappa_\mu$, as is seen from the Fourier transform of (\ref{Cex(n)}):
\begin{equation}\label{Cex(q)}
C_\mu^{(ex)}(k_\nu) = \frac{\sin a\kappa_\mu \cos (k_\nu r_B) + (-1)^\mu \sin ak_\nu \sin (ak_\nu N/2)}{\cos ak_\nu - \cos a\kappa_\mu}.
\end{equation}

We conclude that the kinematic interaction is a weak, but absolutely non-perturbative effect for excitons, especially at large $r_B$. 

\subsection{Kinematic interaction of polaritons}
\label{s-s-KIpolaritons}

\begin{figure}[t]
\centering 
\includegraphics[width=\columnwidth]{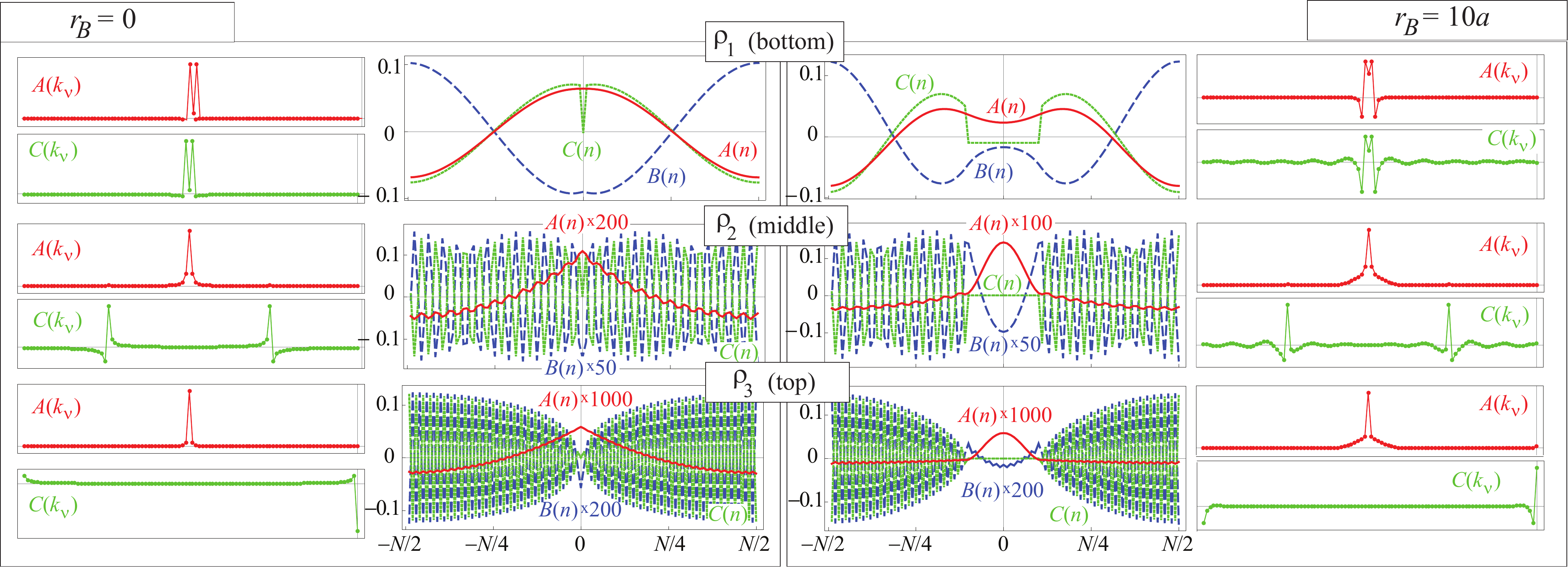}
\caption{Central panels: The photon-photon [$A(n) - \langle A \rangle$; red solid], photon-exciton [$B(n) - \langle B \rangle$; blue dashed] and exciton-exciton [$C(n) - \langle C \rangle$; green dotted] amplitudes  for $r_B=0$ (left) and $r_B = 10a$ (right) for $N=100$ atoms. The first, second and third rows are plotted for the LL-states at $K = 0$, marked, respectively, as $\rho_1, \rho_2$ and $\rho_3$ in figure~\ref{f-spaceholder1}(c), and correspond to the second from the bottom, middle and last states from the LL-band. Upper row demonstrates antibunching of photons for $r_B = 10 a$, while the second and the third rows display bunching for both small and large $r_B$. Smaller side-panels show the corresponding Fourier transforms of the photon-photon and exciton-exciton amplitudes. The Fourier transform of the photon-exciton amplitudes $B(k_\nu)$ behaves exactly as $C(k_\nu)$ (not shown).}
\label{f-compare wave functions}
\end{figure}

Polaritons are hybrid particles consisting of both photons and excitons, with the relative weights of the two components changing as a function of the state index. Near the exciton-photon resonance, in the so-called strong coupling region, the admixture is half-to-half, while out of resonance the upper polariton becomes pure photon-like, and the lower polariton becomes pure exciton-like [see figure \ref{f-spaceholder1}(b)]. Therefore, the scattering properties of a given polaritonic state are determined by (i) its excitonic and photonic content, and (ii) the relative strength of exciton-photon coupling and the interaction.

In order to examine the polariton-polariton kinematic interaction, we solve numerically the Schr\"odinger equation  (\ref{amplitudes_nm})  for the amplitudes $X_\rho = \{ A_\rho, B_\rho, C_\rho\}$ in real space representation. Figure~\ref{f-compare wave functions} shows $X_\rho(n) - \langle X_\rho \rangle$ for three states [marked as $\rho_1$, $\rho_2$ and $\rho_3$ in figure \ref{f-spaceholder1}(c)] of the lower-lower (LL) polariton band [here $\langle X_\rho  \rangle = \sum_n X_\rho(n)/N$], and for two values of the blockade radius, $r_B = 0$ and $r_B = 10a$. For low-energy states, which belong to the strong coupling region (top panels) all three amplitudes have oscillating character, behaving approximately as $\propto\cos[2\pi \rho / (Na)]$, with $C(n)$-amplitudes being suppressed within the forbidden volume $|n| \leq r_B/a$. For large $r_B$ the cut-off in the excitonic amplitude results in a visible depletion of the two-photon amplitude in the Rydberg sphere region: The photons are strongly ``attached" to repelling excitons, and demonstrate antibunching. Looking at the amplitudes in the space of wave vectors $k_\nu$ (smaller panels) one can see that the profiles are dominated by $k_\nu$-states with $\nu \approx \rho$; for $r_B = 10a$ the states are wave packets with a small admixture of wave vectors from the whole Brillouin zone.

\begin{figure}[t]
\centering
\includegraphics[width=0.7\columnwidth]{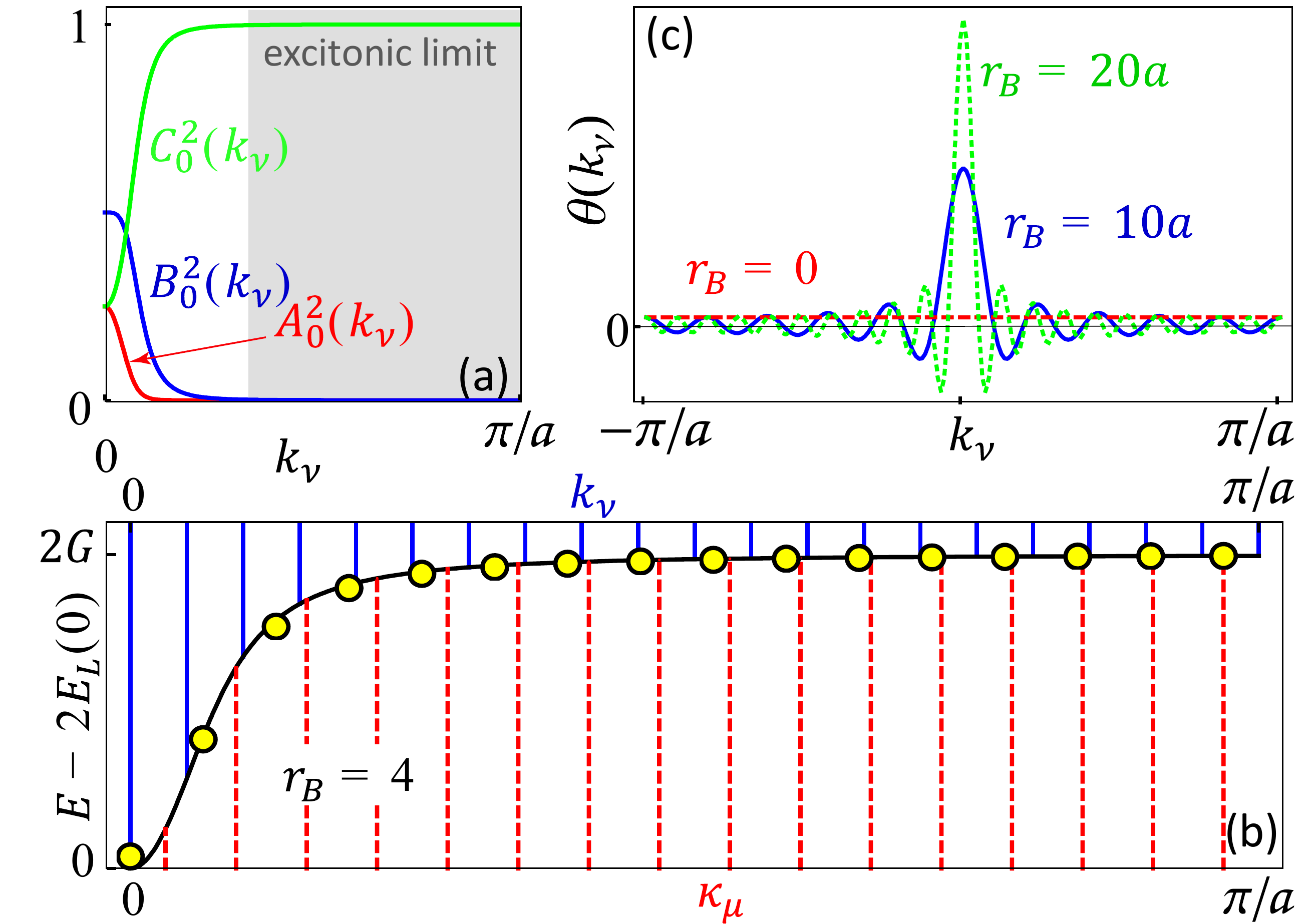}
\caption{(Color online) (a) Photon-photon [$A_0^2(k_\nu)$; red], photon-exciton [$B_0^2(k_\nu)$; blue] and exciton-exciton [$C_0^2(k_\nu)$; green] weights in the absence of interactions. Outside of the strong coupling region ($k_\nu>k_{\rm SC}$, gray shaded area) the total wave function is mostly excitonic. (b) Exact two-polariton energies (yellow circles), compared to the energies of non-interacting lower polaritons $2E_L(k_\nu)$ (black curve) and to the positions of the two wave vector sets $k_\nu$ (blue solid grid) and $\kappa_\mu$ (red dashed grid); here we have chosen $N = 40$ for better visibility. (c) The Fourier transform of the $\theta$-function, $\theta(k_\nu)$, which enters the kernel $F_{\nu\nu'}$ of equation (\ref{F(nu,nu')}), for $r_B = 0$, $10a$ and $20a$ ($N=100$). The maximum magnitude is $[2(r_B/a)+1]/N$.}
\label{f-Lambda}
\end{figure}

At higher energies, i.e.~out of the strong coupling region (larger $\rho$), the situation dramatically changes, as shown in the middle and lower panels of figure~\ref{f-compare wave functions}. The lower panel corresponds to the last state in the LL-band (state index $\rho_3 = N/2-r_B/a$). The middle panel corresponds to the state with the intermediate state index $\rho_2 = (N/2-r_B/a)/2$. The energy of this latter state is close to the band top due to dramatic increase of the lower polariton density of states at higher energies. Most strikingly, the $A$-amplitude loses its oscillatory character and shows well-defined cusps, both in real (near $n=0$) and wave vector (near $k_\nu = 0$) spaces. The larger $r_B$ is, the more dramatic are the deformations. In other words, in spite of the underlying repulsion, photons demonstrate bunching. Although the photonic component of two-polariton states is reduced at larger energies, the bunching increases with $\rho$. The relative magnitude of the photon-photon [$A_0^2(k_\nu)$], photon-exciton [$B_0^2(k_\nu)$] and exciton-exciton [$C_0^2(k_\nu)$] weights for non-interacting polaritons is illustrated in figure~\ref{f-Lambda}(a) [their values were obtained from (\ref{Schroedinger-2}) with $S_\rho(k_\nu) \equiv 0$]. 

An insight into the changes in the polaritonic wave functions can be obtained from viewing two interacting polaritons as composed of two subsystems: A non-interacting one, consisting of photon-photon and photon-exciton states, and an interacting one, consisting of exciton-exciton states. The non-interacting subsystem is described by the quantum numbers~(\ref{k_nu}), the interacting one by those of~(\ref{k renormalized}). The coupling between these subsystems intermixes the two corresponding wave vector sets. In the strong coupling region (lowest energies, small $\rho$) the coupling of excitons to photons dominates over exciton-exciton interaction, and the polaritons are better described by $k_\nu$. With the increase of the state number, polaritons enter the exciton-like regime and are therefore better described by $\kappa_\mu$. Accordingly, the quantum number $\rho$ makes a smooth transition from $\nu$-numbers to $\mu$-numbers. The effect of this crossover for the LL-band can be modelled by introducing a set of effective polaritonic wave vectors:
\begin{equation}
k_\rho^{\rm eff} = \frac{Na - 2r_B - a}{Na - 2r_B} \frac{\pi(\rho - 1)}{Na/2 - r_B -a}, \hskip 1cm \rho = 1, ... N/2-r_B/a.
\end{equation}
For $\rho = 1$, $k_{\rm min}^{\rm eff} = 0$ as for free states, while for $\rho = N/2-(r_B/a)$, $k_{\rm max}^{\rm eff} = \pi[N - 2(r_B/a) -1]/(Na-2r_B)$ as for the last bare exciton-exciton state. Figure~\ref{f-Lambda}(b) shows the exact two-polariton energies calculated numerically (yellow circles), plotted as a function of the effective vectors $k^{\rm eff}_\rho$. Their positions nicely match the analytical estimate $E_{LL}(k_\nu \to k_\rho^{\rm eff})$ for $N=40$, $r_B = 4a$. In \cite{D2-paper} we checked that for $r_B = 0$ the analytical results are essentially exact. The thin vertical lines indicate positions of the ``original" (solid blue) and ``renormalized" (red dashed) wave vector sets. The calculated points indeed exhibit a gradual shift from the first to the second set, demonstrating the crucial role of the wave vector mismatch in the kinematic interaction.

It is convenient to describe the polariton-polariton kinematic interaction in the non-interacting basis $k_\nu$, as this basis corresponds to observable quantities (photons at the exit of the cavity). Due to coupling between interacting and non-interacting subsystems, each $\rho$-state can be viewed as a wave packet formed of specific $k_\nu$ states. To see explicitly the structure of the wave packets and the coupling between the two wave vector sets, we introduce the two-particle operators $\alpha_n^\dag$, $\beta_n^\dag$ and $\gamma_n^\dag$, which describe, respectively, creation of two photons, one photon and one exciton, and two excitons separated by $n$ lattice sites. The two-particle wave function takes the form $\ket{\Psi} = \sum_s \left[ A(s)\ket{\alpha_s} + B(s)\ket{\beta_s} + C(s)\ket{\gamma_s}\right]$. The Hamiltonian we consider is

\begin{equation}\label{H-nu-mu}
\begin{array}{l}
\fl \tilde H_{AB}= \sum\limits_{n,m}\left[2E_p(n-m)\alpha_n^\dagger\alpha_m+\left( E_p(n-m) + E_e(n-m)\right)\beta_n^\dagger \beta_m\right] + G\sqrt{2} \sum_n \left[\alpha_n^\dagger \beta_n+\beta_n^\dagger \alpha_n\right],\\

\\

\fl \tilde H^{(KI)}_{C} =  \sum\limits_{n,m} (1-\theta(n))2E_e(n-m)\gamma_n^\dagger\gamma_m, \hskip 1cm

\tilde H^{(KI)}_{AB-C} = G\sqrt{2}\sum\limits_n (1-\theta(n))\left[\gamma_n^\dagger \beta_n+\beta_n^\dagger \gamma_n\right],\\
\end{array}
\end{equation}
with the last term describing the coupling between the interacting (C) and non-interacting (AB) subsystems. The resulting \Sch equation is identical to the Fourier transform of~(\ref{Schroedinger-2}).

The interaction-free Hamiltonian $\tilde{H}_{AB}$, which describes the ``photon-photon $\bigcup$ photon-exciton" subspace, is diagonalized as
\begin{equation}
\tilde{H}_{AB}= \sum\limits_{i=L,U} \sum\limits_{\nu} E_\nu^{(p,i)} {\xi_{\nu}^{(i)}}^\dagger \xi_{\nu}^{(i)}, \hskip 1cm {\xi_{\nu}^{(i)}}^\dagger = X_\nu^{(i,\alpha)} \alpha_\nu^\dagger + X_\nu^{(i,\beta)} \beta_\nu^\dagger.
\end{equation}
Here and below $i=(L,U)$ is the polaritonic branch index, $\nu$ is the free-state wave index from (\ref{k_nu}), and
\begin{equation}\label{AB-polaritons}
E^{(p,i)}_\nu = E^{(p)}_\nu+\frac{E^{(e)}_\nu + E^{(p)}_\nu \pm \sqrt{\left( E^{(p)}_\nu-E^{(e)}_\nu \right)^2 + 8 G^2} }{2},
\end{equation}
with $E^{(p)}_\nu \equiv E_p(k_\nu)$,  $E^{(e)}_\nu \equiv E_e(k_\nu)$. The energies $E^{(p,i = \{ L,U \})}_\nu$ (with $i=L$ corresponding to ``$-$", and $i=U$ to ``$+$" in the right-hand side) are constructed as sums of energies of one photon and one exciton-polariton with the coupling constant $\sqrt{2}G$, taken at the same wave vector $k_\nu$. They are solutions of the first two lines of equations (\ref{Schroedinger-2}) with $C \equiv 0$. The photon-photon and photon-exciton amplitudes are, respectively,
\begin{equation}\label{X}
\fl X^{(i,\alpha)}_{\nu} = \sqrt{\frac{\left(E^{(p,i)}_\nu - E^{(p)}_\nu - E^{(e)}_\nu\right)^2}{2G^2 + \left(E^{(p,i)}_\nu - E^{(p)}_\nu - E^{(e)}_\nu \right)^2}}, \hskip 1cm X^{(i,\beta)}_{\nu} = \sqrt{1 - \left( X_\nu^{(i,\alpha)} \right)^2}.
\end{equation}

The two-exciton part of the Hamiltonian is diagonalized as
\begin{equation}
\tilde H^{(KI)}_{C} =  \sum\limits_{\mu} E_{\mu}^{(ex)}\chi^\dagger_\mu\chi_\mu, \hskip 1cm \chi_\mu^\dag = \sum\limits_{s =-N/2+1}^{N/2} g_s(\mu)\gamma_s^\dagger,
\end{equation}
with $E_\mu^{(ex)}$ defined in equations (\ref{Eex}), and the state index $\mu$ as in (\ref{k renormalized}). Finally, the interaction Hamiltonian $\tilde{H}^{(KI)}_{AB-C}$ written in terms of $\xi$- and $\chi$-operators reads
\begin{equation}
\tilde H^{(KI)}_{AB-C} = \frac{G}{\sqrt{N[N-2(r_B/a)]}}\sum\limits_{i\nu\mu} \Lambda_{\nu\mu}X^{(i,\beta)}_{\nu}\left(\chi_\mu^\dagger \xi_{\nu}^{(i)} + {\xi_{\nu}^{(i)}}^\dagger \chi_\mu\right),
\end{equation}
with the coefficients $\Lambda_{\nu\mu}$ providing the coupling between the two wave vector sets are

\begin{equation}\label{Lambda}
\Lambda_{\nu\mu} = \frac{\cos\left( \frac{\pi\nu(2r_B+a)}{Na} + \frac{\pi|\mu|a}{Na-2r_B} \right)}{2\sin\left( \frac{\pi\nu}{N} + \frac{\pi|\mu|a}{Na-2r_B} \right)} - \frac{\cos\left( \frac{\pi\nu(2r_B+a)}{Na} - \frac{\pi|\mu|a}{Na-2r_B} \right)}{2\sin\left( \frac{\pi\nu}{N} - \frac{\pi|\mu|a}{Na-2r_B} \right)}.
\end{equation}

At this point, we have expressed the two-polariton problem in the joint basis set consisting of the states $\ket{\chi_\mu}$ (interacting excitons) and $\ket{\xi_{\nu}^{(i=L,U)}}$ (virtual polaritons, representing two-photon and photon-exciton states). We express the wave function as $\ket{\Psi}=\sum_{i\nu}p^{(i)}_\nu\ket{\xi^{(i)}_\nu}+\sum_\mu e_\mu\ket{\chi_\mu}$, and solve the \Sch equation with the Hamiltonian $\tilde H_{AB} + \tilde H^{(KI)}_{C} + \tilde H^{(KI)}_{AB-C}$. The amplitudes $p^{(i)}_\nu$ and $e_\mu$ obey the equations:

\begin{equation}\label{Schroedinger-5}
\begin{array}{l}
\left(E - E^{(p,i)}_\nu\right)p^{(i)}_\nu = \frac{G X^{(i,\beta)}_{\nu}}{\sqrt{N[N-2(r_B/a)]}}\sum\limits_\mu \Lambda_{\nu\mu}e_\mu,\\

\\

\left(E - E_{\mu}^{(ex)} \right)e_\mu =  \frac{G }{\sqrt{N[N-2(r_B/a)]}}\sum\limits_{i\nu} X^{(i,\beta)}_{\nu}\Lambda_{\nu\mu} p^{(i)}_\nu.\\
\end{array}
\end{equation}

Being interested in the $\nu$-representation, we exclude the amplitudes $e_\mu$ from (\ref{Schroedinger-5}). We approximate $E_{\mu}^{(ex)} \approx 2E_0$, which, in the absence of the dynamical interaction, holds as long as $t \ll G$, and obtain a closed system of equations for $ p^{(i)}_\nu$:
\begin{equation}
\left(E - E^{(p,i)}_\nu\right) p^{(i)}_\nu = \frac{G^2 X^{(i,\beta)}_{\nu}}{2N(E-2E_0)}\sum\limits_{i'\nu'} F_{\nu\nu'} X^{(i',\beta)}_{\nu'} p^{(i')}_{\nu'}.
\end{equation}

In this equation, the kernel
\begin{equation}\label{F(nu,nu')}
F_{\nu\nu'}=N\left(\delta_{\nu,\nu'}+\delta_{\nu,-\nu'}\right) - \theta\left[\frac{2\pi}{N}(\nu-\nu')\right] - \theta\left[\frac{2\pi}{N}(\nu+\nu')\right]
\end{equation}
describes the mixing between different $k_\nu$-components -- the formation of the wave packets in the non-interacting subsystem via its coupling to interacting excitons. The first two terms in equation (\ref{F(nu,nu')}) describe the wave vector conserving scattering, and the $\theta$-terms describe the effect produced by the Rydberg sphere. Figure \ref{f-Lambda}(c) shows $\theta(\nu)$ for $r_B = 0$, $10a$ and $20a$. Clearly, larger blockade radii enhance the role of low-$k_\nu$ components in the wave packets.

\subsection{Two-photon bunching}
\label{s-s-bunching}

The equations derived in Section \ref{s-s-KIpolaritons} allow us to quantify the bunching of photons. The amplitude for two photons separated by $n$ lattice sites is related to the amplitudes $p^{(i)}_\nu$ as
\begin{equation}
A(n)=\bra{\alpha_n}\Psi\rangle=\frac{1}{\sqrt{N}}\sum\limits_{i=L,U}\sum\limits_{\nu} p^{(i)}_\nu X^{(i,\alpha)}_{\nu}e^{-\frac{2\pi i\nu n}{N}}.
\end{equation}

 Therefore, $A(0)=\sum_{i\nu} p^{(i)}_\nu X^{(i,\alpha)}_{\nu}/\sqrt{N}$ results from a collective effect of $p$-amplitudes that add up with a vanishing phase; large-separation amplitudes are instead averaged out by the oscillating exponentials. Using equation~(\ref{Schroedinger-5}) we relate $A(0)$ for state $\rho$ to the excitonic amplitudes $e_\mu$:
\begin{equation}
\label{bun-2}
A_\rho(0)=\frac{G}{N\sqrt{N-2(r_B/a)}} \sum\limits_{i=L,U}\sum\limits_{\nu} \frac{X^{(i,\alpha)}_{\nu} X^{(i,\beta)}_{\nu}}{(E_\rho - E^{(p,i)}_\nu)} \sum\limits_{\mu}\Lambda_{\nu\mu} e_{\mu}.
\end{equation}

Finally, for a given $\rho$, the amplitudes $e_\mu \equiv e_\mu^{(\rho)}$ are related to $C_\rho(s)$ via $2e_\mu^{(\rho)} = \sum_s g_s(\mu)C_\rho(s)$ [this results from the equality $C_\rho(s) = \bra{\gamma_s^{(\rho)}}\Psi_\rho \rangle = \sum\limits_\mu e_\mu^{(\rho)} g_s(\mu)$ following from the representation of $\ket{\Psi_\rho}$ via $A,B,C$- and $p,e$-amplitudes]. We then make the following approximation, valid for larger $\rho$, i.e.~for the energies outside of the strong coupling region, where the bunching occurs: In the equation $e_\mu^{(\rho)} = 0.5\sum_s g_s(\mu)C_\rho(s)$, we approximate the polaritonic $C_\rho$-amplitudes by the closest in energy (i.e. with $\mu = \rho - 1/2$) bare exciton-exciton amplitudes $C_{\mu}^{(ex)}$ (\ref{Cex(n)}) multiplied by a normalization coefficient $X^{(\gamma)}_\rho \lesssim 1$, which accounts for the presence of finite exciton-photon and photon-photon excitation in the total wave function of $\rho$-th eigenstate. Using the orthogonality of the $g$-functions, we obtain the final form for $A(0)$:
\begin{equation}
\label{bun-4}
A_\rho(0) \approx \frac{G X^{(\gamma)}_\rho}{\sqrt{2N[N-2(r_B/a)]}} \sum\limits_{i = L,U} \sum\limits_{\nu} \frac{X^{(i,\alpha)}_{\nu} X^{(i,\beta)}_{\nu}}{(E_\rho - E^{(p,i)}_\nu)} \Lambda_{\nu,\rho-\frac{1}{2}}.
\end{equation}

\begin{figure}[t]
\centering
\includegraphics[width=0.9\columnwidth]{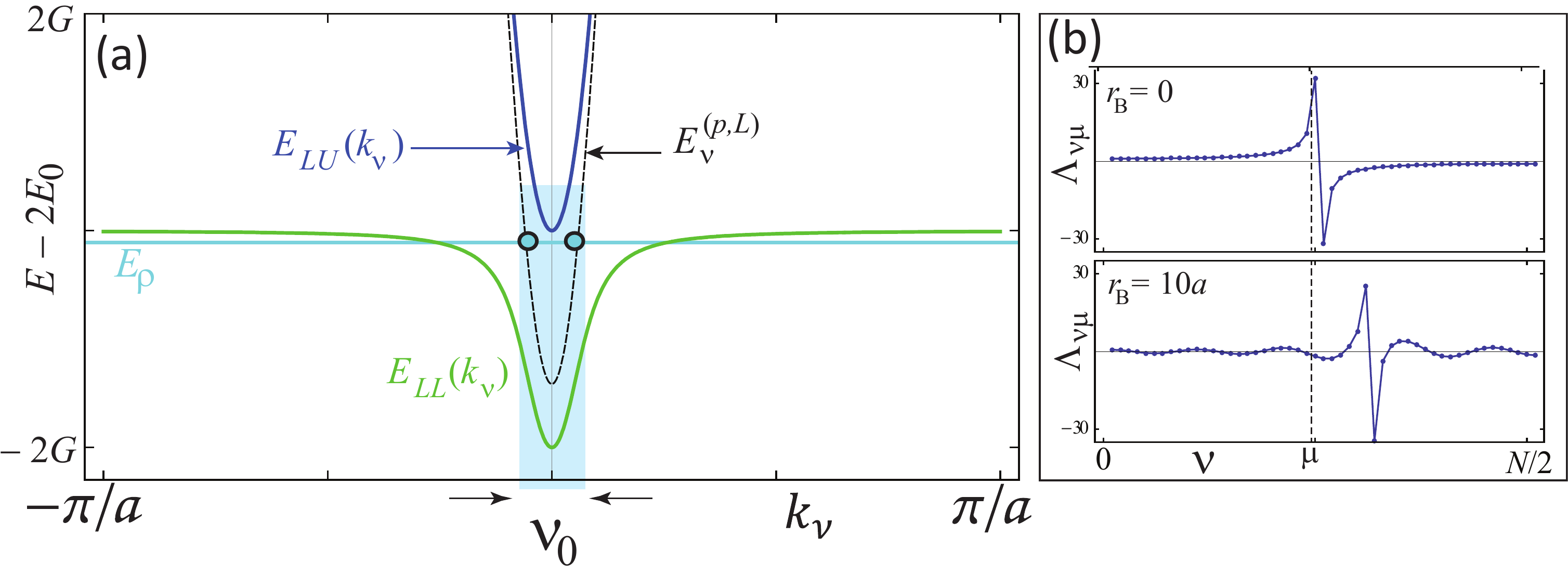}
\caption{(Color online) (a) Formation of the wave packets. The shaded region shows low-$k_\nu$ states selected by the resonance between a state $E_\rho$ and virtual scattering states $E_\nu^{(p,L)}$. (b) Mixing coefficients $\Lambda_{\nu\mu}$ as function of the integer index $\nu$, at $\mu = N/4 - 1/2$ for $r_B = 0$ and $r_B = 10a$.}
\label{f-bunching-explain}
\end{figure}

This equation allows us to explaine the criteria that determine the bunching. First, due to the factor $X^{(\gamma)}_{\rho}$, bunching appears only for states with visible exciton-exciton content, which rules out the states above the top of LL-band. Second, the resonant denominator \footnote{
Note that the quasi-pole appearing in this sum is not a real pole. Due to the mismatch in the quantum numbers, in the framework of a discrete model the energy states $E_\rho$ always fall between the neighboring states of ideal system.}
$(E_{\rho} - E^{(p,i)}_\nu)^{-1}$ selects -- out of the total set of $\nu$-states -- only those states with $\nu_0$ such that $E^{(p,i)}_{\nu_0} \approx E_{\rho}$. Note that $E^{(p,i)}_{\nu_0}$ in (\ref{AB-polaritons})  are {\it not} non-interacting polaritonic states. Instead, they are the eigenstates of the non-interacting subsystem, and can be interpreted as providing a virtual scattering channel through which excitons interact. Indeed, we observed that the bunching is established only when $E_{\rho}$ is resonant with the $E^{(p,L)}_\nu$-band; instead, when $E_{\rho_0} < \min\left\{ E^{(p,i)}_\nu  \right\} = E^{(p,L)}_{\nu=0}$, the photons antibunch (for large $r_B$), or remain unperturbed (for small $r_B$).

Therefore, the denominator selects a $\nu_0$-set illustrated in figure \ref{f-bunching-explain}(a): The final two-photon state (with the account of kinematic interaction) is a wave packet formed of $A(k_{\nu_0})$. Due to the strong dispersion of $E^{(p,L)}_{\nu}$ (shown by black dashed line in the figure), only the low-$\nu$ states are resonant with $E_\rho$ belonging to LL-band. This is why the two-photon wave packets $A(k_\nu)$ shown in figure \ref{f-compare wave functions} have a peak centered around $k_\nu = 0$, and vanishing contribution from high-$k_\nu$ states. The real-space Fourier transform of this cusp-like shape is again a peak-like (bunching) profile [see the high-$\rho$ $A(n)$-profiles in figure \ref{f-compare wave functions}].

Apart of the normalization $1/\sqrt{N - 2(r_B/a)}$, the dependence of $A_\rho(0)$ on the blockade radius $r_B$  is contained in the coupling coefficients $\Lambda_{\nu,\rho-1/2}$. The latter have resonant shape, with a maximum at $\nu = \pm(\rho-1/2)N/[N-2(r_B/a)]$ [see figure \ref{f-bunching-explain}(b); these coefficients are responsible for the $C(k_\nu)$ profiles of figure \ref{f-compare wave functions}]. In the next section we show that a larger $r_B$ results in the enhancement of the bunching strength accompanied by the narrowing of the frequency window for this effect.


\subsection{Numerical results}

\begin{figure}[t]
\centering
\includegraphics[width=0.7\columnwidth]{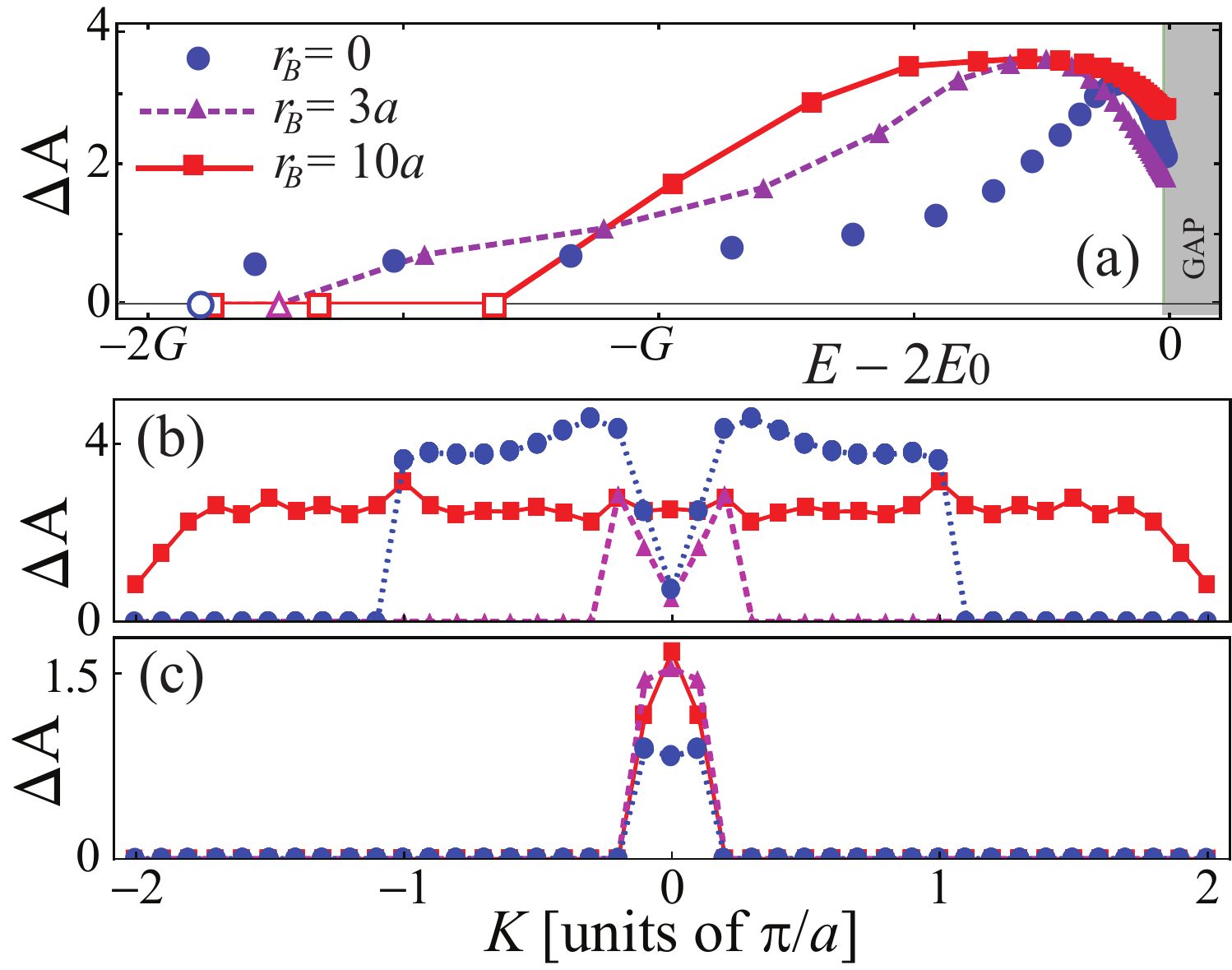}
\caption{(Color online) Quantifying bunching. (a) Figure of merit $\Delta A$ in the LL-band for $r_B = 0, 3a$, and $10a$. Empty symbols indicate the states showing two-photon antibunching; the gray shaded region is the gap between the $LL$-band and the $LU$-band (not shown). (b-c) Figure of merit $\Delta A$ as a function of the total wave vector $K_{\nu'}$ for $r_B = 0, 3a$, and $10a$, for the highest (b) and lowest (c) energy state among all LL-states showing bunching.}
\label{f-bunching}
\end{figure}

We define bunching as the situation when $|A_\rho(n)|$ has a global maximum at $n=0$. The antibunching is defined as a global minimum of $|A_\rho(n)|$ at $n=0$. We quantify the bunching magnitude by a relative figure of merit:

\begin{equation}
\hspace{-2cm}\Delta A(E_\rho) = \left\{
\begin{array}{lll}\displaystyle
\frac{| A_\rho(n=0)| - \langle |A_\rho(n)| \rangle_n}{\langle |A_\rho(n)| \rangle_n} >0 , & {\rm if~there~is~bunching}\\

0, & {\rm otherwise}\\
\end{array}
\right.
\end{equation}
with $\langle |A_\rho(n)| \rangle_n=\sum_n |A(n)|/N$. Figure~\ref{f-bunching}(a) shows $\Delta A(E_\rho)$ for $r_B = 0$, $3a$ and $10a$ ($N = 100$); empty symbols mark antibunched states. The plot presents results for small positive detuning ($\delta = 0.1 G$), and the gap between LL- and (lower-upper-) LU-bands is highlighted by gray color. The photonic bunching is clearly visible in a wide range of scattering states with a frequency window $\sim G$, which is in the GHz range for typical atomic parameters (i.e., a comparatively broadband effect). With the increase of $r_B$, the bunching strength increases, but the frequency window for the effect shrinks, with antibunching replacing bunching for low-energy states with strong exciton-photon correlation.

Up to now, we considered two-excitation states with $K_{\nu'} = 0$. For applications, the existence of correlations at finite $K_{\nu'}$ is crucial, as they refer to the propagation of the center of mass of two polaritons. These states allow for exactly the same numerical analysis utilised above, with the difference that for them the asymmetric part of the exciton-photon amplitude, $B^A(n)$, is non-zero, and all equations are more cumbersome (see Appendix). In figure~\ref{f-bunching}(b-c) we plot the magnitude of $\Delta A(K_{\nu'})$ for $r_B = 0$, $3a$ and $10a$ for a state chosen in the very top [panel (b)] and at the bottom [panel (c)] of the band of LL-states showing bunching. The plots confirm that the bunching exists for a wide range of total wave vectors and blockade radii.

Other control parameters are the detuning $\delta$, the coupling strength $G$, and the lattice constant $a$. Basically, every choice of these parameters that increases the phase space occupied by the strong coupling region in the first Brillouin zone results in the increase of the bunching strength \cite{D2-paper}. By approximating the small-$k_\nu$ dispersion of the lower polariton by a quadratic dispersion, we can estimate the size of the strong coupling region, $k_{\rm SC}$ (marked in figure~\ref{f-spaceholder1}) by the intersection of this parabola with $E_0$. Then the criterium for stronger bunching reads: $k_{\rm SC} = 2\sqrt{E_0 G}/(c\hbar) \sim \pi/a$. This relation, never satisfied in natural solids, where $ak_{\rm SC} \sim 10^{-4}$, can be easily fulfilled in atomic systems. In particular, larger $G$ and larger $a$ are better for bunching. Positive detuning decreases the strength of interaction between photon and exciton, and effectively has the same effect on the polariton dispersion curve as the reduction of the coupling constant $G$. Consequently, positive detuning results in gradual suppression of bunching strength. On the contrary, negative detuning leads to the increase of the bunching figure of merit $\Delta A$, -- however, the bunching appears for a narrower frequency interval, as the LL-band overlaps with the LU-band and the upper-lower- (UL-) band for higher $\rho$. All these arguments equivalently apply to usual and extended kinematic interaction.

\section{Dynamical interaction and bound states}
\label{s-D}

In this section we study the effect of the dynamical interaction in addition to the hard-core repulsion. First of all we find that for repulsive interactions ($D > 0$), the continuum states show no visible changes up to values of $D / [2(r_B/a) + 1]^3 $ of the order of several $G$: The photonic bunching remains essentially unaltered for repulsive interactions. The reason for that can be traced back to the Fourier transform of (\ref{S(k)}): The term responsible for kinematic interaction (and bunching) dominates over the effect of the dynamical interaction as long as $D(n) C(n) \ll G\sqrt{2} \theta(n) B(n)$ ($G \gg t$ is assumed). This condition is definitely satisfied inside the blockade region, where the maximum of the photonic amplitude develops. The minor variations in the bunching profile we have observed can be attributed to the renormalization of the excitonic energies due to interactions. In contrast, the dynamical interaction can be significant for attractive interaction ($D < 0$), as the latter can produce bound bipolariton states, which may considerably change the spectra, as we show below.\\

We start our analysis from bound biexciton states, as they are precursors for bipolaritons. With no kinematic interaction and hopping, the dispersion equation for biexcitons is as simple as
\begin{equation}\label{biexciton}
(E - 2E_0) C_{nm} = D_{nm} C_{nm}.
\end{equation}
This brings us to an oversimplified, but yet very instructive, reference picture of biexcitonic levels that split off the continuum. The number of split levels equals the number of interacting neighbors taken into account [for which $D(n-m)$ is not considered as vanishing]. For long-range dipole-dipole and van der Waals interaction, the $n$-th state is split, respectively, by $D/n^{3}$ and $D/n^{6}$, where $D = D_{nm}(|n-m| = 1)$. The amount of broadening present in the system determines whether the $n$-th state can be resolved.

Bipolaritonic states can be obtained from the equations (\ref{amplitudes_K(k)}) given in the Appendix, which are the $K \neq 0$ analogues of (\ref{Schroedinger-2}). Neglecting hopping ($t \ll G$) and eliminating $A_K(k)$ and $B_K(k)$, the following integral equation for $C_K(k)$ is obtained (where the $\nu$-subscripts for wave vectors are dropped for shortness):
\begin{equation}\label{C_K(k)}
C_K(k) = \frac{1 + \phi_K(E, k)}{N(E - 2E_0)} \sum\limits_q D(k-q) C_K(q),
\end{equation}
where
\begin{equation}
\phi_K(E, k) = \frac{2G^2[E - E_{p}(K/2+k) - E_p(K/2-k)]}{\Delta_K(k)}.
\end{equation}

In the nearest neighbor approximation it is possible to derive a closed-form analytical expression for the eigenstates of this equation by separating the variables $k$ and $q$: Indeed, $D(k-q) = 2D \cos a(k-q) = 2D[\cos ak \cos aq + \sin ak \sin aq]$. The nearest neighbor approximation is routinely used for the usual kinematic interaction with $r_B = 0$ in the context of dipole-dipole interaction. For the extended kinematic interaction, when $r_B \gg a$, this assumption is clearly insufficient, and the account of long-range interactions to all orders seems more appropriate. However, we can combine the nearest-neighbor polaritonic model with the intuition coming from the examination of the biexciton levels with long-range interactions, described by (\ref{biexciton}): It suggests that each ``next neighbor" will add one more bipolariton band split from the continuum. We found that this model captures all essential physics and proves useful even for larger $r_B$.

Based on this latter picture, we assume that the nearest neighbor approximation holds, and introduce two quantities $\alpha_K = \sum_q C_K(q) \cos aq$ and $\beta_K = \sum_q C_K(q) \sin aq$. Then equation (\ref{C_K(k)}) splits into the following two

\begin{equation}
\begin{array}{c}
(E - 2E_0)\alpha_K = \frac{2D}{N} \left\{ \alpha_K \sum_K [1 + \phi_K(E, k)] \cos^2 ak + \right.\\

\\

\left. + \beta_K \sum_K [1 + \phi_K(E, k)] \sin ak \cos ak \right\},\\

\\

(E - 2E_0)\beta_K = \frac{2D}{N} \left\{ \alpha_K \sum_K [1 + \phi_K(E, k)] \sin ak \cos ak + \right.\\

\\

\left. + \beta_K \sum_K [1 + \phi_K(E, k)] \sin^2 ak \right\}.\\
\end{array}
\end{equation}

This system admits a non-trivial solution provided that its determinant vanishes, which yields:

\begin{equation}\label{E_pp}
\begin{array}{c}
\displaystyle
\left( E - 2E_0 - D \biggl[ 1 + \frac{2}{N} \sum\limits_k \phi_K(E, k) \sin^2 ak \biggr] \right) \times\\

\\

\displaystyle \times \left( E - 2E_0 - D \biggl[ 1 + \frac{2}{N} \sum\limits_k \phi_K(E,k) \cos^2 ak \biggr] \right) =\\

\\

= \displaystyle \left(\frac{2D}{N} \sum\limits_k \phi_K(E,k) \sin ak \cos ak \right)^2.\\
\end{array}
\end{equation}

\begin{figure}[t]
\centering
\includegraphics[width=0.7\columnwidth]{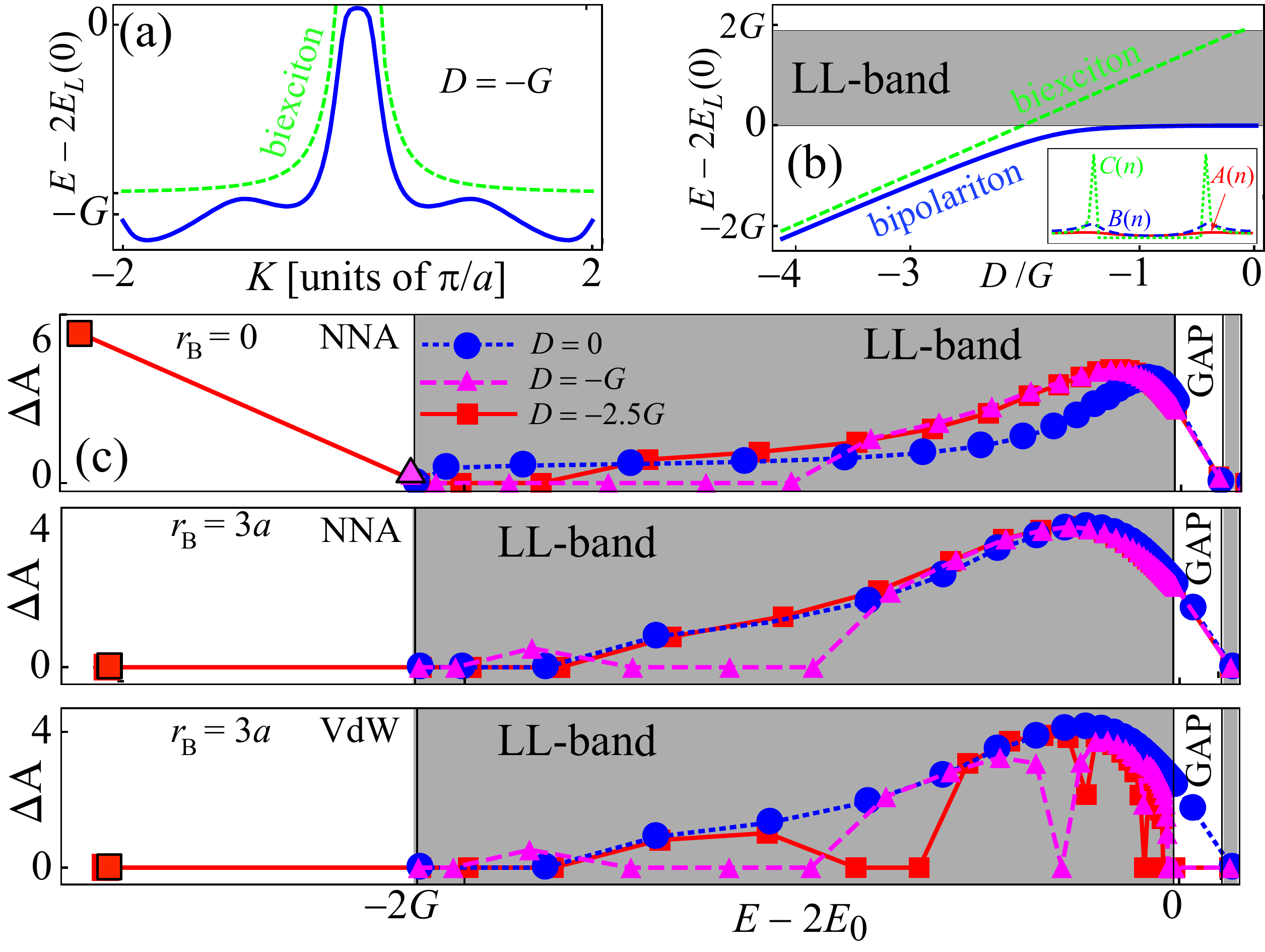}
\caption{(Color online) Bound states for attractive dynamical interaction. (a) The energy of the bipolariton as a function of $K$ (zero detuning, $D = -G$). (b) Relative energy of the bipolariton (blue line) with respect to the LL-continuum (gray band) and to the biexciton energy (green dashed line); $K=0$, same $\delta$ and $G$. (c) Bunching strength $\Delta A$ in the LL-band for $D = 0$ (blue dotted), $D = -G$ (weak dynamical interaction; magenta dashed line) and $D = -2.5 G$ (strong interaction; red solid line). LL-band is shown by a gray background. The bipolariton is the lowest state on the white background below the LL-band, which appears at $D = -2.5 G$. Top: $r_B = 0$, same plot for VdW and NNA interactions. Middle and lowest panel: $r_B = 3a$, NNA and VdW interaction, respectively.}
\label{f-D attractive}
\end{figure}

Each bracket in the left-hand side of this equation reminds of the solution of the eigenequation for excitons (\ref{biexciton}) with an interaction constant $D$ renormalized by the light-matter coupling. Only the second of these two brackets, however, yields a symmetric solution. The first one corresponds to an asymmetric $C$-amplitude, which is unphysical, and therefore has to be omitted. The remaining equation has one solution, which we denote by $E_{bP}(K)$, and which is located below the LL-continuum band -- a dynamical bipolariton. The dependence $E_{bP}(K)$ is shown in figure \ref{f-D attractive}(a) for $D = -G$, $r_B = 0$.

The dispersion equation (\ref{E_pp}) takes the most simple form for $K=0$, as then $\phi_{K=0}(E,k)$ is a symmetric function of $k$, and the right-hand side of (\ref{E_pp}) vanishes. Then:
\begin{equation}\label{E_b}
\fl E = 2E_0 + D \left[ 1 + \frac{4G^2}{N} \sum\limits_k 
\frac{E - 2E_p(k)}{\Delta(E,k)} \cos^2 ak \right], \hskip 1cm E = E_{bP}(K=0).
\end{equation}

The position of the bipolariton level with respect to LL-continuum and the resulting properties of the wave functions depend on the ratio between $D$ and $G$. Figure~\ref{f-D attractive}(b) shows $E_{bP}(K = 0)$ as a function of $D$ at fixed $G$ (solid blue line). It allows us to distinguish between two regimes:\\

{\noindent \it (i) Limit of strong dynamical interaction} ($|D| > 1.5 G$)

The bipolariton level is well-split from the continuum, and asymptotically tends to the biexciton energy, $E_{bE}^{(1)} = 2E_0 - |D|$ shown by green dashed line in figures~\ref{f-D attractive}(a,b). The wave function of this single split state has dominating excitonic character with two sharp peaks at the minimal possible separation $2r_B + a$ [figure \ref{f-D attractive}(b), inset; $D = -4G$].\\

{\noindent \it (ii) Limit of weak dynamical interaction} ($|D| < 1.5 G$)

In this regime the bipolariton level is basically drown in the continuum. In this case most of the continuum states have a biexcitonic feature superimposed on their own structure. This feature is especially pronounced for the eigenstates close in energy to $E_{bE}^{(1)}$.\\

Now we use this simple picture to interpret the numerical results for polariton-polariton attraction for realistic long-range interactions in the presence of non-zero $r_B$. We use two model interaction potentials: ``nearest neighbor" approximation (NNA), where the dynamical interaction vanishes for all separations larger than $2r_B+a$:
\begin{equation}
D_{nm}^{(NNA)} = D \delta(|n|-2r_B/a-1),
\end{equation}
and  the long-range van der Waals (VdW) potential:
\begin{equation}\label{eq:VdW}
D_{nm}^{(VdW)} = D \theta(n) \left[\frac{2(r_B/a)+1}{n}\right]^6,
\end{equation}
where the factor $[2(r_B/a)+1]^6$ is introduced for better matching between these two potentials: both give the same value, $D$, when $n = 2(r_B/a)+1$.

The data shown in figure \ref{f-D attractive}(c) represent the bunching strength, $\Delta A$, calculated for zero (blue dotted), moderate [magenta dashed, regime (ii)] and strong [red solid, regime (i)] dynamical interaction. Let us start with the NNA approximation (upper and middle panels). In complete accordance with our simple model, for large $D$ one state splits off the lower polariton continuum for all three panels (red square). For $r_B = 0$ this state shows strong bunching [figure \ref{f-D attractive}(c), upper panel] -- as in this lowest eigenstate photons are strongly coupled to excitons, which tend to attract each other. However, for large $r_B$ the same argument leads to the suppression of bunching in the lowest-energy split state: Photons follow the excitons, which, in spite of the attraction, are now separated by the distance $2 r_B + a$ [figure \ref{f-D attractive}(b), inset]. Figure~\ref{f-D attractive}(c) also illustrates the difference between the weak- and strong-$D$ regimes for the continuum states. In the limit (i) the continuum states remain mainly untouched by dynamical interaction: The bipolariton state is off-resonance and does not influence the continuum. In contrast, in limit (ii), the bipolariton interacts with the continuum, which suppresses bunching.

Finally, we can discuss the role of long-range terms for dynamical interaction of polaritons [lowest panel of figure \ref{f-D attractive}(c)]. By comparing the lowest and the middle panels,  we see that taking these interaction terms into account does not change the tendency to two-photon bunching, except for a suppression of bunching in several continuum states for large $r_B$; for $r_B = 0$ the plots for all $D$ remain exactly the same as for NNA. We explain this suppression of bunching by a presence of LU-bound states, which split from the LU-continuum, enter the LL-continuum and superimpose a biexcitonic feature (resulting in the suppression of bunching at large $r_B$) onto the resonant LL-wave functions. With further increase of $D$, the second LL-bound state comes out of the LL-continuum and produces a second split bound bipolariton state, etc.

We conclude this section by pointing out that these bound states are two-level intracavity analogs of the bound states described in \cite{Buchler} for cavity-free three-level Rydberg atoms. Here, we have presented a unified approach, which not only describes these states, but also shows how they influence the photonic bunching in the continuum. In summary, the dynamical interaction, whether attractive or repulsive, does not eliminate bunching in the LL-continuum. Attractive forces can produce a bound state (or, at larger $D$ and long-range interaction, a series of bound states) below LL-band. For small $r_B$ this narrow spectral feature exhibits strong bunching in the photonic component. Bound bipolariton states that are split from the LU-band can suppress bunching for some of the LL-continuum states at large $r_B$.

\section{Gap states}
\label{s-gap}

The polaritonic spectra depend on the detuning $\delta = E_p(0) - E_0$ between cavity mode and excitonic resonance. If the detuning is larger than the hopping constant $t$, there is a gap between the LL and the LU-bands for non-interacting polaritons, see figure~\ref{f-spaceholder1}. In this section we discuss the states, which can form inside this gap as a result of polariton-polariton interactions. These states are bound states, and appear as narrow resonances with vanishingly (for small decoherence rates) weak coupling to the continuum, and therefore can be of considerable interest for engineering highly controllable two-photon states.

\begin{figure}[t]
\centering
\includegraphics[width=0.9\columnwidth]{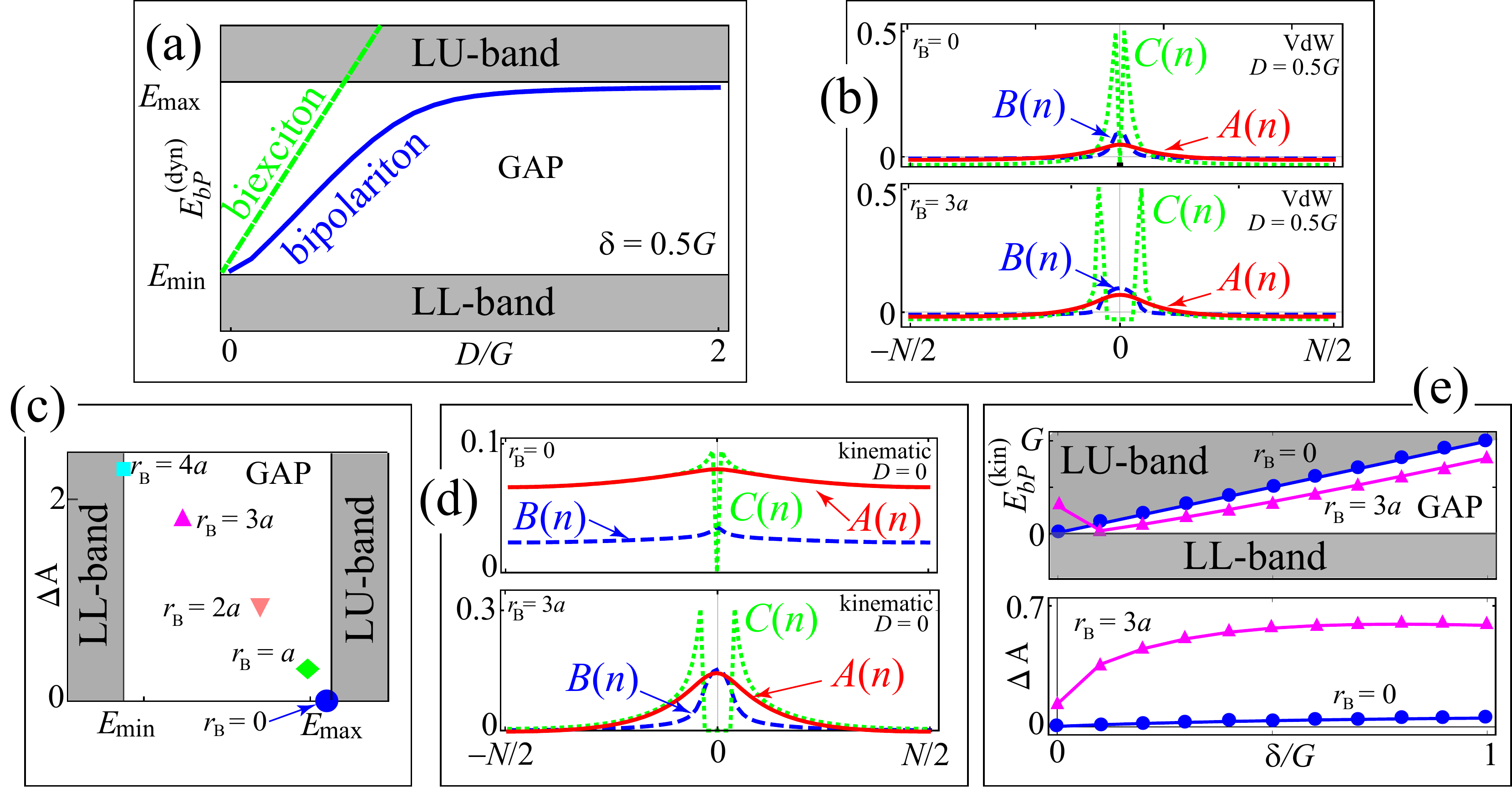}
\caption{(Color online) In-gap bound states for repulsive dynamical and kinematic interactions. (a) $D>0$, position of the bipolariton and biexciton levels in the gap (white space stretched between two gray continua showing LL- and LU-bands).  (b) $D = 0.5G$, wave functions of the bipolariton for VdW interaction and $r_B=0$, $3a$. (c-d) $D=0$, gap bipolariton for $r_B$ ranging from $0$ to $4a$ and associated wave functions for $r_B=0$ and $3a$.  (e) $D=0$, location of the bipolariton level in the gap (upper panel) and height of the bunching figure of merit $\Delta A$ (lower panel) as a function of detuning for $r_B=0$, $3a$.}
\label{f-D repulsive}
\end{figure}

One type of bound states can be formed by the sole dynamical interaction. In fact, (\ref{E_b}) is valid for both signs of $D$. For positive detuning and repulsive interaction, a bound state can form in the gap between the LL- and LU-bands. This is an example of a pair bound by repulsive forces in gapped spectra \cite{Winkler,Piil,Wong}. Figure~\ref{f-D repulsive}(a) shows $E_{bP}(K = 0)$ as a function of $D/G$ at fixed $\delta = 0.5 G$, and figure \ref{f-D repulsive}(b) demonstrates typical wave functions for $D = 0.5 G$ -- the in-gap analogs of figure~\ref{f-D attractive}(b). These wave functions exhibit strong bunching, which, in parallel with bunching in the continuum states, can be used for inducing correlations between photons. The plot is obtained for long-range van der Waals interaction. For $D > G$ the bipolariton approaches the LU-band and may mix with it. In fact, by tuning $D$ one may vary the position of this single bunching feature, either placing it in the gap (``bunching on"), or placing it into the continuum (``bunching off"). Alternatively, if $D$ is fixed, one may vary the detuning, as this will also change the location of the bound state.

Another type of a gap bound state can appear even with $D = 0$, owing only to the kinematic interaction \cite{D2-paper}. It turns out that, as a result of the kinematic repulsion, the lowest LU-state can split from the LU-band and drop into the gap region. Figure~\ref{f-D repulsive}(c) shows its location in the gap for $\delta = 0.1 G$ for several values of $r_B$. The splitting from the LU-band grows with the increase of $r_B$, as a result of the enhancement of the kinematic interaction. The bound character of this state is confirmed by the shape of its wave functions, see an example in figure \ref{f-D repulsive}(d). This state is a {\it kinematic bipolariton}, similar to the kinematic biexciton appearing in organic crystals with two molecules in a unit cell \cite{Basko}. As we noted in \cite{D2-paper}, the kinematic biexciton overlaps with the continuum band, and can be easily destroyed by any coupling mechanism to the latter (due to disorder or phonons). In contrast, the kinematic bipolariton discussed here is located in the gap, and is stable against decoherence. Figure \ref{f-D repulsive}(e) shows the location of the kinematic bipolariton within the gap as a function of detuning (upper panel), and the corresponding two-photon bunching (lower panel) for $r_B = 0$ and $r_B = 3a$. While for $r_B = 0$ the biexciton is actually attached to the LU-continuum for all detunings, for $r_B = 3a$ the bipolariton lays deep in the gap for $\delta \sim G$, and exhibit strong bunching. The latter drops to zero when $\delta \to 0$. Clearly, large blockade radii are beneficial for the formation of kinematic bipolaritons with narrow and strong bunching features.

\section{Discussion and concluding remarks}
\label{s-conclusions}

We have shown that 1D cavity polaritons exhibit a range of nonlinearities, originating from both dipole forces and the hard-core character of excitations in the atomic subsystem. The nonlinearities are typically manifested as bunching of photons. The bunching induced by kinematic interaction appears in the GHz frequency range in the continuum states, while dipolar forces may produce a narrow state exhibiting bunching that is split from the continuum. A similar type of split-off state can appear due to the kinematic interaction alone, provided that there is a gap in the polariton spectrum.

In this proof-of-the-principle paper we disregarded all sources of broadening. This is substantiated by the fact that our lattice-trapped two-level configuration excludes many sources of decoherence. This is immediately clear for collisional and Doppler losses. Moreover, atomic resonant absorbtion, which often acts as a bottle-neck for atom-light applications, is absent on principle in our configuration. This is due to the combined effect of the ordering of atoms, and overlapping volumes for atomic and photonic subsystems. The translational invariance leads to formation of coherent atomic modes (excitons) characterized by a wave vector. As a result, instead of irreversible decay of one-atom modes to photonic continuum, Rabi oscillations between one collective excitonic mode with a given $k$ and one photonic mode with the same wave vector are established. In other words, radiative decay of a single atom is replaced by coherent evolution between a photon and a collective atomic excitation -- i.e. between two extended excitations occupying the same quantization volume.

Therefore, in order for this model to hold, the collective light-matter coupling constant $G$ must exceed all broadenings yet present in the system. This requires $(i)$ careful preparation of a Mott insulator state in the optical lattice, and $(ii)$ long living time for the photons within the waveguide or cavity. Furthermore, our model assumes coupling between one excitonic resonance and one photonic mode. This requires, firstly, using, e.g., narrow fibers, so that one of the lowest photonic modes of the fiber is in resonance with the transition: $E_0 \approx E_p(0) = c \hbar k_\perp$. For atomic transitions in the range of hundreds of THz, the resonance condition with the lowest cavity mode is achieved at the radius $R$ being a fraction of a micron. We further require that the splitting between different atomic levels must exceed $G = (d/R)\sqrt{2 E_0/a}$. These conditions, which can be readily met for low-lying transitions in atoms \cite{D2-paper}, impose the following range of restrictions on the principal quantum numbers $n$ of Rydberg atoms: When $n$ is too large, the separation between levels becomes small, such that the photon couples to a set of excitonic resonances. In this case we expect similar, but not identical, physics to that presented in this work. On the other hand, larger $n$ are needed in order to achieve blockade radii that largely exceed $a$. In three-level Rydberg atoms under EIT conditions the blockade radius is defined as $r_B^{(EIT)} = (C_6/\gamma_{EIT})^{1/6}$, where $C_6$ is the coefficient entering the van der Waals potential, and $\gamma_{EIT}$ is the width of the EIT window with typical value of several MHz: If an atom in the ground state approaches the excited atom by a distance shorter than $2r_B$, its energy levels are detuned off the EIT regime, and the atom cannot interact with light. In our system it is more appropriate to define $r_B = (C_6 / 2G)^{1/6}a$ in order to take the second atom out of the strong coupling region. The large energy scale of $G$ may make it difficult to achieve large $r_B$ values. 

Similar physics to that presented here is expected to arise as well in a two-dimensional (2D) atomic ensemble placed in a planar microcavity, where photons also possess dispersion with a cut-off frequency. Wedge-shaped planar cavities allow for the control of detuning, thus making possible additional tuning of the bunching. Differences with the discussion above will definitely appear for the bound states, as their number will double in 2D \cite{Vektaris}.

A very interesting question is whether the bunching of the continuum states described here can survive in the out-of-lattice intracavity geometry, i.e. omitting the ordering requirement, but preserving the requirement of similar quantization volumes for atoms and light. In principle, the formation of polaritons in disordered atomic ensembles is possible in some solid materials \cite{lidzey}, while the argument of the reduced quantization volume and the $k_\nu - \kappa_\mu$ mismatch remains valid. For sufficiently dense trapped Rydberg gases, formation of polaritons should be possible \cite{gas polaritons}, as long as the broadening induced by Rayleigh scattering is smaller than $G$. Note that the gap kinematic bipolariton (which exhibits a strong bunching) is expected to be found in the continuum as well. At finite Rayleigh scattering, light will interact with an incoherent set of spatially distributed atomic transitions with a finite linewidth, rather than with a coherent plane-wave-like quasiparticle with a finite bandwidth. The analysis of the free-gas geometry will be the subject of future work.

It would be very exciting to observe the effects described in this work not only in atomic ensembles, but also in solids. As we discussed in \cite{D2-paper}, the bunching strength depends on the relative size of the strong coupling region, and therefore natural semiconductors are not good candidates. On the other hand, extended kinematic interactions with $r_B \gg a$ may become realistic in solids, in view of recent experiments \cite{CuO}, where Wannier-Mott Rydberg excitons with $n$ up to 20 were observed. A detailed analysis of losses will be crucial in this case.


\section*{ACKNOWLEDGEMENTS}
We acknowledge support by the ERC-St Grant ColdSIM (No.~307688). Additional support from ANR via BLUESHIELD, UdS via Labex NIE and IdEX, RYSQ.\\


\appendix
\section{\Sch equation for arbitrary total wave vectors}

The two-polariton wave function is given by (\ref{Psi}) with the constraint (\ref{constraint}). We solve the \Sch equation for the Hamiltonian (\ref{Hamiltonian}). We eliminate the basis states $\ket{P_n P_m}$ by multiplying the corresponding amplitudes by $[1-\theta(n-m)]$, and symmetrize the obtained equations. The result is:

\begin{equation}
\label{amplitudes_nm}
\begin{array}{l}
\fl EA_{nm} =  \sum\limits_s[E_p(n-s) A_{sm} + E_p(m-s) A_{ns}] + G\sqrt{2} B_{nm}^S,\\

\\

\fl EB_{nm}^S =  E_0B_{nm}^S + \frac{1}{2} \sum\limits_s[E_p(n-s) (B_{sm}^S + B_{sm}^A) + E_p(m-s)(B_{sn}^S+B_{sn}^A)] + \\

\\

\hskip -1cm + G\sqrt{2} (A_{nm} + C_{nm}) +\\

\\

\hskip -1cm + \frac{t}{2}\left[(B_{nm-1}^S+B_{nm-1}^A+B_{nm+1}^S+B_{nm+1}^A)+(B_{n-1m}^S-B_{n-1m}^A+B_{n+1m}^S-B_{n+1m}^A)\right],\\

\\

\fl EB_{nm}^A =  E_0 B_{nm}^A + \frac{1}{2} \sum\limits_s[E_p(n-s) (B_{sm}^S + B_{sm}^A) - E_p(m-s) (B_{sn}^S+B_{sn}^A)] + \\

\\

\hskip -1cm + \frac{t}{2}\left[(B_{nm-1}^S+B_{nm-1}^A+B_{nm+1}^S+B_{nm+1}^A)-(B_{n-1m}^S-B_{n-1m}^A+B_{n+1m}^S-B_{n+1m}^A)\right],\\

\\

\fl EC_{nm} =  2E_0C_{nm} + (1-\theta(n-m)) G\sqrt{2} B_{nm}^S  + D(n-m) C_{nm} +\\

\\

\hskip -1cm + t(C_{nm-1}+C_{nm+1}+C_{n+1m}+C_{n-1m}) (1-\theta(n-m)).\\
\end{array}
\end{equation}

We Fourier transform these equations, and introduce the total and relative wave vectors, $K_{\nu^{'}} = q_{\nu_1}+q_{\nu_2}$ and $k_{\nu} = (q_{\nu_1} - q_{\nu_2})/2$. The total wave vector $K_{\nu^{'}}$ is the quantum number describing the two-excitation spectra. We rewrite the amplitudes as $A(q_{\nu_1}, q_{\nu_2}) \to A_{K_{\nu^{'}}}(k_\nu)$, etc., and introduce the two-particle energies of bare excitations:
\begin{equation}\label{Eij}
\fl E_{ij}^{K_{\nu^{'}}}(k_\nu) = E_i(K_{\nu^{'}}/2 + k_\nu) + E_j(K_{\nu^{'}}/2 - k_\nu) \equiv E_i(q_{\nu_1}) + E_j(q_{\nu_2}), \hskip 1cm i,j \in e,p.
\end{equation}

For a given $K_{\nu^{'}}$, the resulting equations for the amplitudes in the $k_\nu$-space are:

\begin{equation}\label{amplitudes_K(k)}
\begin{array}{l}
\fl EA_{K_{\nu^{'}}}(k_\nu) = E_{pp}^{K_{\nu^{'}}}(k_\nu) A_{K_{\nu^{'}}}(k_\nu) + G\sqrt{2} B^S_{K_{\nu^{'}}}(k_\nu),\\

\\

\fl EB^S_{K_{\nu^{'}}}(k_{\nu}) = \frac{1}{2} [E_{pe}^{K_{\nu^{'}}}(k_{\nu})+E_{pe}^{K_{\nu^{'}}}(-k_{\nu})] B^S_{K_{\nu^{'}}}(k_{\nu}) + \frac{1}{2} [E_{pe}^{K_{\nu^{'}}}(k_{\nu})-E_{pe}^{K_{\nu^{'}}}(-k_{\nu})] B^A_{K_{\nu^{'}}}(k_{\nu}) + \\

\\

\hskip -0.5cm + G\sqrt{2} (A_{K_{\nu^{'}}}(k_{\nu}) + C_{K_{\nu^{'}}}(k_{\nu})),\\

\\

\fl EB^A_{K_{\nu^{'}}}(k_{\nu}) = \frac{1}{2} [E_{pe}^{K_{\nu^{'}}}k_{\nu})+E_{pe}^{K_{\nu^{'}}}(-k_{\nu})] B^A_{K_{\nu^{'}}}(k_{\nu}) + \frac{1}{2} [E_{pe}^{K_{\nu^{'}}}(k_{\nu})-E_{pe}^{K_{\nu^{'}}}(-k_{\nu})] B^S_{K_{\nu^{'}}}(k_{\nu}),\\

\\

\fl EC_{K_{\nu^{'}}}(k_{\nu}) = 2E_{ee}^{K_{\nu^{'}}}(k_{\nu}) C_{K_{\nu^{'}}}(k_{\nu}) + G\sqrt{2} B^S_{K_{\nu^{'}}}(k_{\nu}) - \frac{G\sqrt{2}}{N} \sum\limits_{q_{\nu}} \theta(k_{\nu}-q_{\nu}) B^S_{K_{\nu^{'}}}(q_{\nu}) - \\

\\

\hskip -0.5cm - \frac{4t}{N} \sum\limits_{q_{\nu}} \theta(k_{\nu}-q_{\nu}) C^S_{K_{\nu^{'}}}(q_{\nu}) \cos \frac{aK_{\nu^{'}}}{2}\cos aq_{\nu}+ \frac{1}{N} \sum\limits_{q_{\nu}} D(k_{\nu}-q_{\nu}) C_{K_{\nu^{'}}}(q_{\nu}).\\
\end{array}
\end{equation}

In the forth equation, the last term is the dynamical interaction, and the ones with $\theta$-function are the contribution of the kinematic interaction. When $K_{\nu^{'}}=0$, the amplitude $B^A$ vanishes, the energies $E_{ij}^{K_{\nu^{'}}}(k_\nu) \to E_i(k_\nu) + E_j(k_\nu)$, and all the equations greatly simplify [see (\ref{Schroedinger-2})].

%
%
%
%

\end{document}